\documentclass{aa}
\usepackage{astron,times,graphics}
\begin{document}
\thesaurus{06(08.16.4; 08.03.4; 08.09.2 IRAS15194--5115; 08.12.1; 13.09.6)} 
\title{The carbon star IRAS\,15194-5115: Circumstellar CO radio and 
FIR rotational lines\thanks{Based on observations with ISO, an ESA 
project with instruments funded by ESA Member States (especially the 
PI countries: France, Germany, the Netherlands and the United Kingdom) 
and with the participation of ISAS and NASA.  }}

\author{N. Ryde\inst{1} \and F.L. Sch\"oier\inst{2} \and H.
Olofsson\inst{2}
}

\institute{Uppsala Astronomical Observatory, Box 515, SE-751 20
Uppsala, Sweden
\and Stockholm Observatory, SE-133 36 Saltsj\"obaden, Sweden}

\offprints{N.~Ryde (ryde@astro.uu.se)}
\date{Received ; accepted}
\maketitle
\begin{abstract}
We have modelled radio and far-infrared rotational lines of $^{12}$CO 
and $^{13}$CO obtained from the circumstellar envelope of the 
infrared-bright carbon star \object{IRAS\,15194-5115}.  Eleven 
rotational lines between $J$=1--0 and 21--20 and nine rotational lines 
between $J$=1--0 and 22--21 in the ground vibrational states of 
$^{12}$CO and $^{13}$CO, respectively, provide the observational 
constraints.  A model of the circumstellar envelope with a constant 
mass-loss rate (1$\times$10$^{-5}$\,\mbox{M$_\odot$\,yr$^{-1}$}) and
 $^{12}$CO/$^{13}$CO-ratio (5.5) is consistent with our 
observed FIR and radio data, suggesting that the wind characteristics 
have not changed significantly over the past few thousand years.  
Thus, \object{IRAS\,15194-5115} appears to be a highly evolved 
AGB-star, but the carbon star properties combined with the inferred 
low $^{12}\mbox{C}/^{13}\mbox{C}$-ratio make the evolutionary status 
of this star uncertain.  It may have been a J-star for which the 
$^{12}\mbox{C}/^{13}\mbox{C}$-ratio has remained low, or it may be a 
star of 5 to 8 solar masses, which has recently become a carbon star 
due to quenching of hot bottom burning.  The dust properties or the 
relative amount of dust in the envelope appear to be different from 
those in the envelope of the well known carbon star 
\object{IRC+10\,216}.

\keywords{stars: AGB and post-AGB -- circumstellar matter --
individual: \object{IRAS 15194-5115} --
late-type -- Infrared: stars}
\end{abstract}

\section{Introduction}

Red giant stars lose large amounts of matter through a massive, but
slow, stellar wind during their late evolution on the Asymptotic Giant
Branch (AGB).  The mass loss results in circumstellar envelopes (CSEs),
several thousand stellar radii in size, consisting of gas, mainly in
molecular form, and dust.  Important information on late stellar
evolution can be obtained from the circumstellar emission (Olofsson 
1996)\nocite{olofsson96}.

We have studied the third brightest (at 12\,\mbox{$\mu$m}) carbon star (C-star)
\object{IRAS\,15194-5115} (or II Lup), which is the
brightest one known in the southern hemisphere, in several rotational
transitions in the ground vibrational states of $^{12}$CO and
$^{13}$CO.  The lowest energy lines, $J$=1--0 etc., lie at radio
wavelengths and probe the outer regions of the CSE. Higher energy lines,
for example, the $J$=14--13 to $J$=22--21 lines presented in this paper,
which lie in the far-infrared (FIR) wavelength range 100--200\,\mbox{$\mu$m}
originate from regions much closer to the star. By combining observations in
these two wavelength ranges we are able to probe different regions of
the CSE, and to compare wind characteristics, such as mass loss rate,
expansion velocity, molecular abundances (including isotope ratios),
over a time span of several thousand years.

\object{IRAS\,15194-5115} was discovered by Epchtein et al.
\nocite{epstein} (1987; the ESO-Valinhos 2.2\,\mbox{$\mu$m} sky
survey) and by IRAS in the far-infrared, and it was identified as a
C-star using optical and infrared spectra (Meadows et al.
1987)\nocite{meadows}.  To understand the late evolutionary phases of C-stars,
in which they undergo intense mass loss, this star provides an object which
is bright enough to make possible an exhaustive spectroscopic
investigation over a broad range of wavelengths.
\object{IRAS\,15194-5115} shows similarities in the infrared to the
well-studied prototype C-star \object{IRC+10\,216} (Meadows et al.
1987\nocite{meadows}; Lopez et al.  1993\nocite{lopez}).  On the basis
of radio observations Nyman et al.  \cite*{nyman93} concluded that
most molecular species, except a few, have circumstellar abundances
comparable to those in \object{IRC+10\,216}.  Also, they found that
the two objects have comparable mass loss rates.  The most striking
difference between these two C-stars is the $^{12}$C/$^{13}$C-ratio,
which appears to be at least five times lower in
\object{IRAS\,15194-5115} (Nyman et al. 1993\nocite{nyman93}).

Here, we present observations of FIR rotational lines
of $^{12}$CO and $^{13}$CO from \object{IRAS\,15194-5115}, as well
as a model interpreting these data and radio rotational lines.

\begin{figure*}
  \resizebox{15cm}{!}{\includegraphics{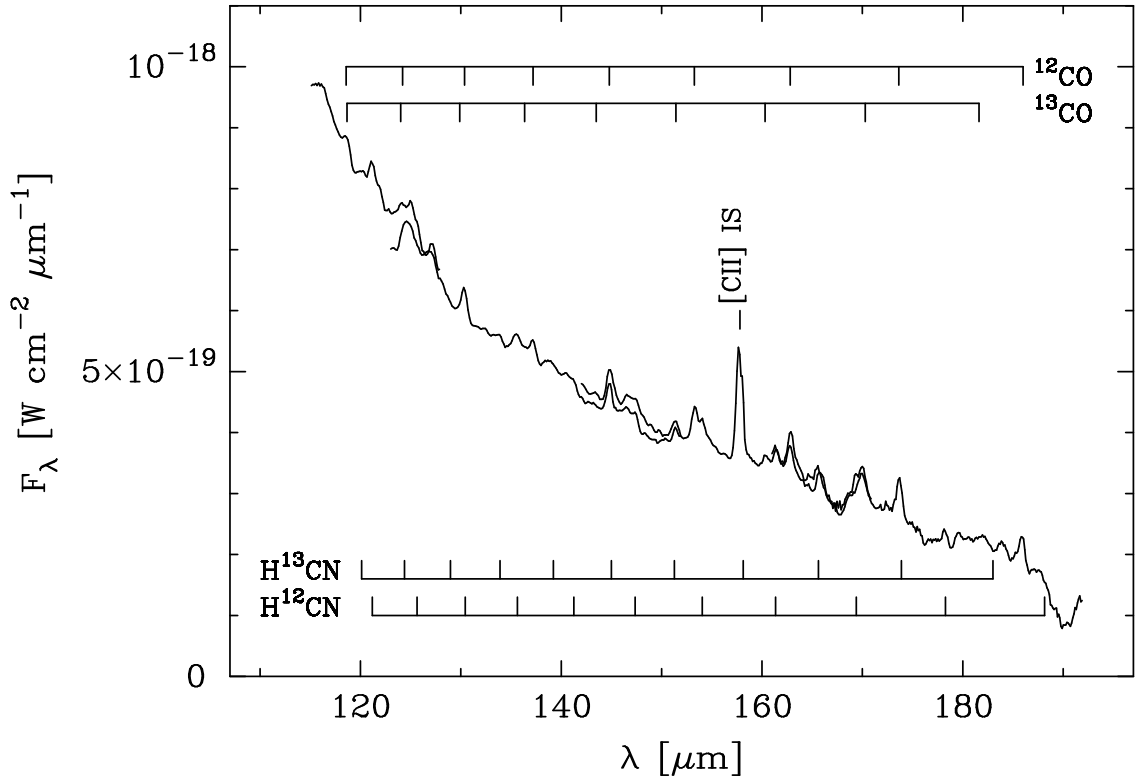}}
  \caption{The FIR spectrum of \object{IRAS\,15194-5115} as observed by the ISO satellite. 
  Fluxes from four detectors are shown without scaling. The overlap regions 
  show the uncertainties, both the relative and the absolute ones, in these 
  ISO-LWS observations.  All rotational transitions within the ground 
  vibrational states of $^{12}$CO, $^{13}$CO, H$^{12}$CN, and H$^{13}$CN 
  in this region are indicated.  Also the interstellar [C{\sc ii}] line is 
  marked.  These emission lines are superimposed on the continuum 
  emitted by the dust around the star}
  \label{FIR}
\end{figure*}

\section{Observations}

We have obtained FIR observations with the Infrared Space Observatory
(Kessler et al. 1996\nocite{kessler}), see Fig.~\ref{FIR}.  
The Long Wavelength Spectrometer (LWS, Clegg et al. 1996\nocite{LWS}), 
with a beam size assumed to be
$70\mbox{$\arcsec$}$, was used in the grating mode (LWS01), covering
the range 90-197\,\mbox{$\mu$m}, and providing a mean spectral
resolution element of $\sim$0.7\,\mbox{$\mu$m} (this corresponds to
${\rm R=\lambda/\Delta \lambda \sim 100-250}$; hence the circumstellar
lines are far from being resolved).  The spectra were sampled at
0.15\,\mbox{$\mu$m}.  The 3.5$^{\mathrm{h}}$ of observations were performed on 19
September 1997 during ISO revolution 674.  The reductions were made
using the most recent pipeline basic reduction package OLP (v.7) and the ISO
Spectral Analysis Package (ISAP v.1.5).  The pipeline processing of
the data, such as wavelength and flux calibration, is described in
Swinyard et al.  \cite*{swinyard}; the combined absolute and systematic 
uncertainties in the fluxes are of the order of
$\pm$30\%.  The accuracy of the wavelength varies and can be as bad as
$\pm$0.1\,\mbox{$\mu$m}. The observed fluxes are measured after scaling 
the four detectors and averaging the two measurements in the overlap 
regions between different LWS detectors. The scaling needed is at 
most a few percent. In Fig.~1 the fluxes from the four detectors are 
shown without any scaling.  Additional uncertainties in the line fluxes, 
sometimes large, are introduced in the continuum subtraction.  A dust 
continuum level is estimated and subtracted over a limited wavelength 
interval, usually $\sim$10$\mu$m.  The errors in the fluxes due to the 
uncertain continuum subtraction are estimated to be around $\pm$20\%. 
A few CO-lines are blended by an unknown amount of H$^{13}$CN, leading to 
somewhat too high measured CO-fluxes in these lines. 

Our final LWS spectrum is dominated by thermal radiation due to the
large quantity of dust around the object.  The dust modelling of the object
by Lopez et al.  \cite*{lopez} shows that for wavelengths longer than
10\,\mbox{$\mu$m} the spectrum is totally dominated by dust emission.
Superimposed on this emission are molecular emission lines, of which we have
identified those due to $^{12}$CO, $^{13}$CO, and H$^{12}$CN (all 
relevant rotational transitions within the ground vibrational states of 
$^{12}$CO, $^{13}$CO, H$^{12}$CN, and H$^{13}$CN are indicated in 
Fig.~\ref{FIR}).  Here we present eight lines between $J$=14--13 to 
21--20 of $^{12}$CO and seven lines between $J$=15--14 and 22--21 of 
$^{13}$CO, Table~1 and Fig.~\ref{iso}.  A similar LWS spectrum of 
\object{IRC+10\,216} has been presented by Cernicharo et al.  
(1996\nocite{cerni}).  The observed fluxes from this source are an 
order of magnitude higher than those towards 
\object{IRAS\,15194-5115}, making the analysis more difficult in our 
case.

\begin{figure*}
  \resizebox{\hsize}{!}{\includegraphics{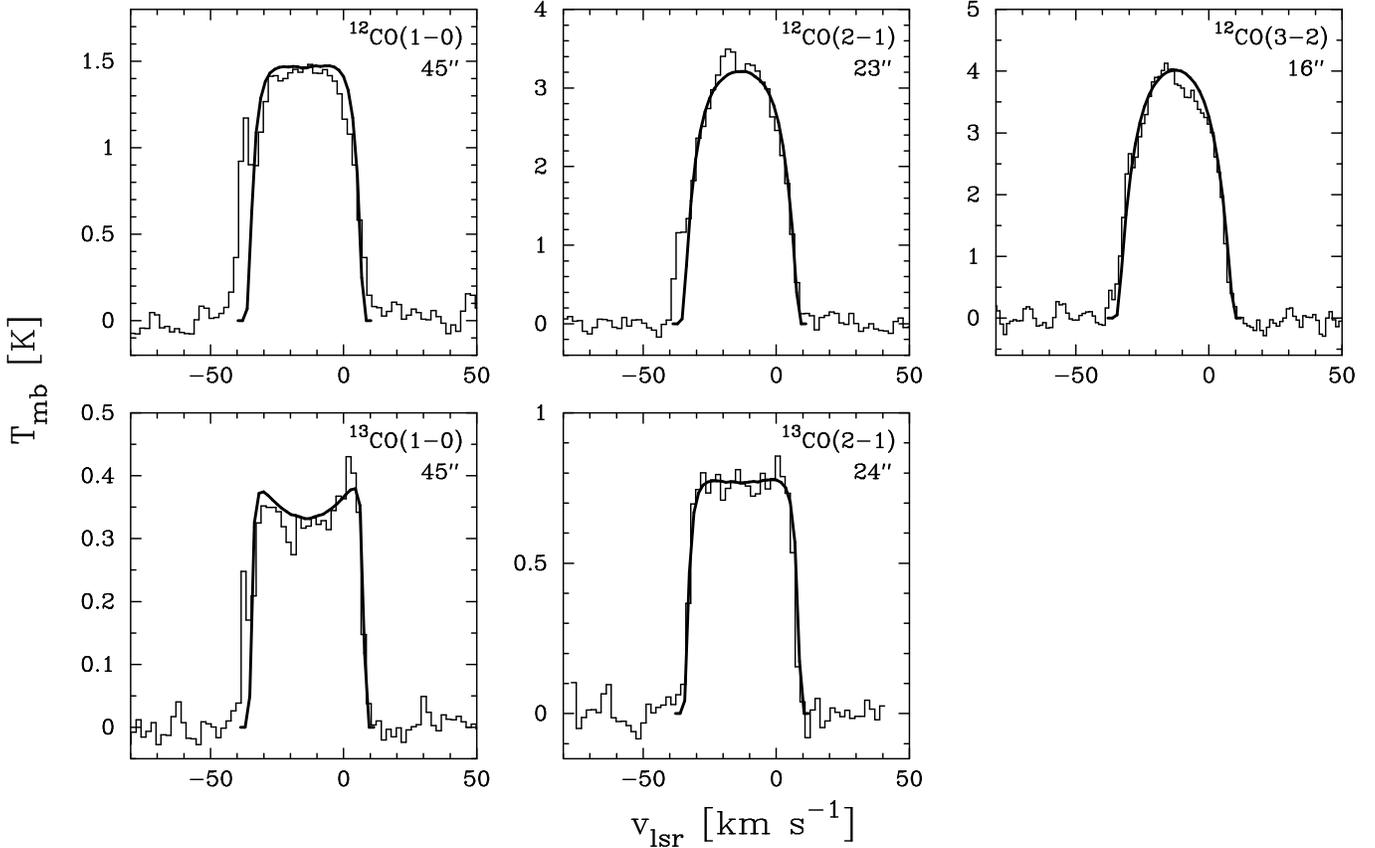}}
  \caption{Millimetre-wave CO spectra (histogram) observed
  towards \object{IRAS\,15194-5115} using SEST, overlaid with the best model
  results (full line) using a mass loss rate of
  1$\times$$10^{-5}$\,\mbox{M$_\odot$\,yr$^{-1}$} and a
  $^{12}\mbox{CO}/^{13}\mbox{CO}$ ratio of 5.5.  The beam size of each
  observation is given}
  \label{radio}
\end{figure*}

\begin{figure*}
  \resizebox{\hsize}{!}{\includegraphics{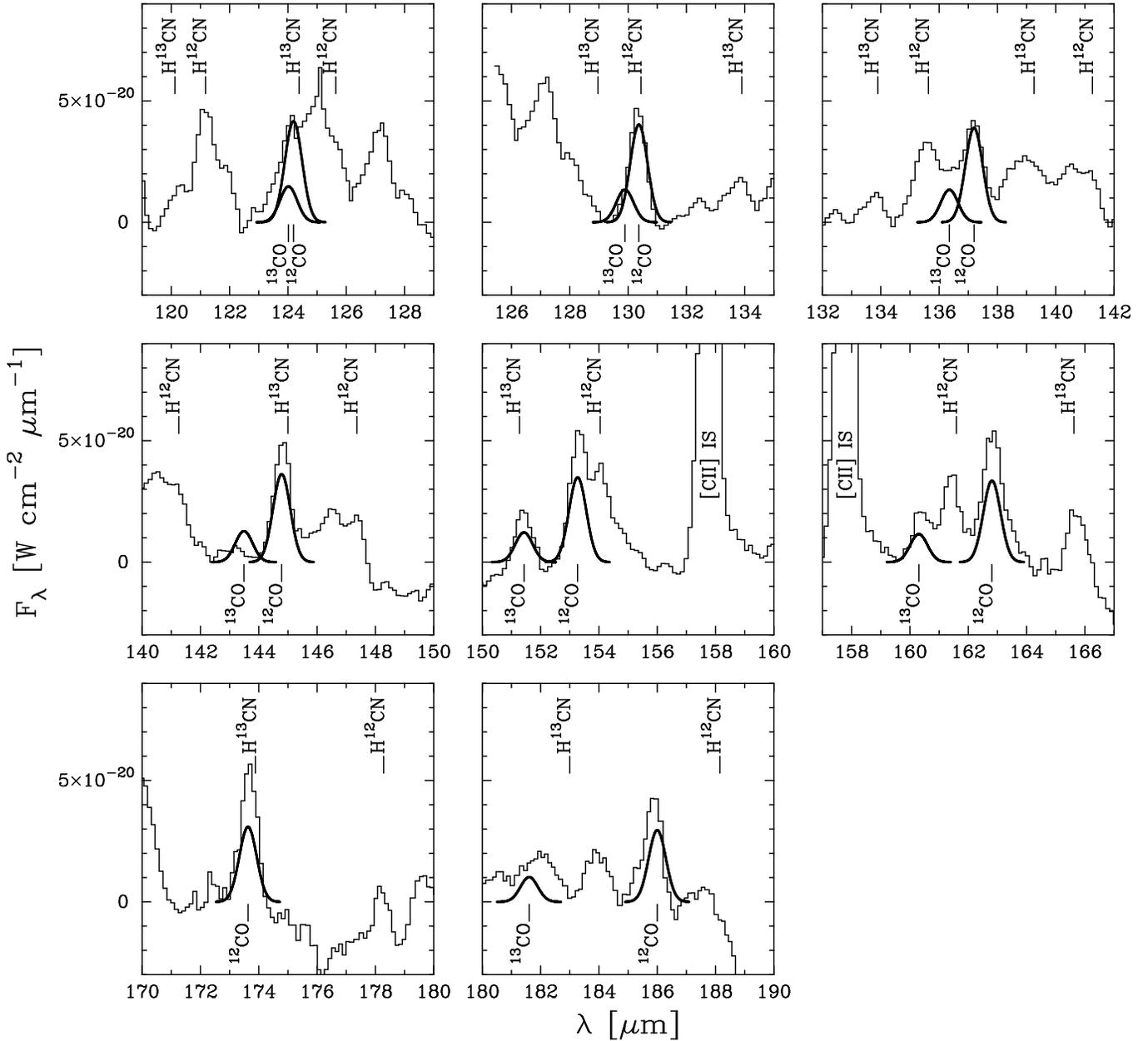}}
  \caption{Selected continuum-subtracted FIR CO spectra (histogram) 
  observed towards \object{IRAS\,15194-5115} using the ISO satellite, 
  overlaid with the best model results (full line) using a mass loss 
  rate of 1$\times$$10^{-5}$\,\mbox{M$_\odot$\,yr$^{-1}$} and a 
  $^{12}\mbox{CO}/^{13}\mbox{CO}$ ratio of 5.5.  For all transitions the 
  beam size and spectral resolution used are $70\arcsec$ and $0.7\mu$m, 
  respectively }
  \label{iso}
\end{figure*}

As a supplement to the ISO data, we have used $^{12}$CO and $^{13}$CO
radio-line observations from SEST [Swedish-ESO Sub-millimetre
Telescope; the $J$=1--0 and 2--1 spectra published in Nyman et al.
(1993\nocite{nyman93}), and a $^{12}$CO(3--2) spectrum (Nyman, private
communication)].  The radio data are presented in Fig.~\ref{radio}.  The errors
in the radio intensities are about $\pm$10\%, and
the resolution is better than 1\,km/s.

\begin{table}
  \caption{Observed and modelled CO FIR rotational lines from
  \object{IRAS\,15194-5115}.
    A colon (:) marks lines with uncertain flux estimates due to a complex continuum around 
    the line around 143.5\mbox{$\mu$m} or due to a blending of 
    two or three lines of $^{12}$CO, $^{13}$CO or HCN}
  \label{tab_iso}
  \begin{tabular}{l l r l l} \hline
  \noalign{\smallskip}
  \multicolumn{1}{c}{$\lambda_\mathrm{obs}$} &
  \multicolumn{1}{c}{$F_\mathrm{obs}$} &
  \multicolumn{1}{c}{Transition} &
  \multicolumn{1}{c}{$\lambda_\mathrm{vac}$} &
  \multicolumn{1}{c}{$F_\mathrm{mod}$} \\
  \multicolumn{1}{c}{[\mbox{$\mu$m}]} &
  \multicolumn{1}{c}{[\mbox{W\,cm$^{-2}$}]} &
   &
  \multicolumn{1}{c}{[\mbox{$\mu$m}]} &
  \multicolumn{1}{c}{[\mbox{W\,cm$^{-2}$}]} \\
  \noalign{\smallskip}
  \hline
  \noalign{\smallskip}
  185.99 & 3.0$\times 10^{-20}$ &        CO(14--13) & 185.999 & 2.2$\times
10^{-20}$ \\
  181.61 & 1.3$\times 10^{-20}$ & $^{13}$CO(15--14) & 181.608 & 7.6$\times
10^{-21}$ \\
  173.66 & 3.8$\times 10^{-20}$ &        CO(15--14) & 173.631 & 2.3$\times
10^{-20}$ \\
  162.85 & 4.2$\times 10^{-20}$ &        CO(16--15) & 162.811 & 2.5$\times
10^{-20}$ \\
  160.54 & 1.5$\times 10^{-20}$ & $^{13}$CO(17--16) & 160.305 & 8.6$\times
10^{-21}$ \\
  153.38 & 4.0$\times 10^{-20}$ &        CO(17--16) & 153.267 & 2.6$\times
10^{-20}$ \\
  151.37 & 1.2$\times 10^{-20}$ & $^{13}$CO(18--17) & 151.431 & 9.1$\times
10^{-21}$ \\
  144.80 & 3.8$\times 10^{-20}$ &        CO(18--17) & 144.784 & 2.7$\times
10^{-20}$ \\
  143.40 & 6.2$\times 10^{-21}:$ & $^{13}$CO(19--18) & 143.493 & 9.5$\times
10^{-21}$ \\
  137.12 & 3.5$\times 10^{-20}$ &        CO(19--18) & 137.196 & 2.9$\times
10^{-20}$ \\
  136.35 & 1.5$\times 10^{-20}$ & $^{13}$CO(20--19) & 136.352 & 1.0$\times
10^{-20}$ \\
  130.28 & 1.9$\times 10^{-20}:$ &        CO(20--19) & 130.369 & 3.0$\times
10^{-20}$ \\
  129.83 & 6.2$\times 10^{-21}:$ & $^{13}$CO(21--20) & 129.891 & 1.0$\times
10^{-20}$ \\
  124.10 & 2.8$\times 10^{-20}:$ &        CO(21--20) & 124.193 & 3.1$\times
10^{-20}$ \\  
  124.10 & 9.3$\times 10^{-21}:$ & $^{13}$CO(22--21) & 124.019 & 1.1$\times
10^{-20}$ \\
  \noalign{\smallskip}
  \hline
  \end{tabular}
\end{table}

\section{The model}

In order to model the circumstellar emission lines we have used a
numerical simulation code, based on the Monte Carlo method, to solve
the non-LTE radiative transfer of $^{12}$CO and $^{13}$CO taking into
account 30 rotational levels for each of the ground and first
vibrational states (see Sch\"oier, PhD thesis in prep., for details).
Using 40 energy levels does not change the temperature structure 
and the line intensities noticeably. For example, in
the innermost regions, cooling by CO excited to higher levels 
than $J$=20--25 are unimportant.  Crosas \& Menten (1997)\nocite{crosas97} 
also found that around 30 levels are sufficient to treat the CO 
cooling in CSEs properly.  The decline in CO cooling in the inner 
regions of the CSE,  as demonstrated in Fig.~4, is due to the fact that the CO 
lines become optically thick.


The transition probabilities and energy levels are taken from Chandra et al.
\cite*{chandra}, and the rotational (CO-H$_2$) collisional rate
coefficients are from Flower \& Launay \cite*{flower} (they are
extrapolated for $J$$>$11 and for temperatures higher than $250~\mbox{K}$).
The derived intensities of the FIR lines are not very sensitive to the
accuracy of this extrapolation since they are formed in LTE (see below).
Collisional transitions between vibrational levels are not important
due to the low density and the short radiative lifetimes.

The kinetic gas temperature is calculated in a self-consistent way,
using the derived level populations, by solving the energy balance
equation (e.g. Goldreich \& Scoville 1976)\nocite{goldreich76},

\begin{equation}
\frac{dT}{dr} = (2-2\gamma)\frac{T}{r} + \frac{\gamma-1}{n_{\rm
               H_2}kv_\infty} (H-C),
\label{ebal}
\end{equation}

\noindent where $\gamma$ is the adiabatic index, $n_{\rm H_2}$ the number 
density of molecular hydrogen which is the most abundant species in the 
CSE, $v_\infty$ the terminal gas expansion velocity, $k$ the Boltzmann 
constant, $H$ the total heating rate per unit volume, and $C$ the 
total cooling rate per unit volume.

The first term on the right-hand-side in the energy balance equation 
is the cooling due to the adiabatic expansion of the gas.  The 
adiabatic index, i.e.  the ratio of the specific heats 
$\gamma=\mathrm{c_p/c_v}$, can be assumed to be 5/3 since H$_2$ can be 
treated as a `mono-atomic gas'.  This is true as long as the 
corrections on $\gamma$ due to rotational excitation of H$_2$ are 
small.  These corrections are important for temperatures greater than 
approximately $300~\mbox{K}$, i.e.  for the innermost regions where 
$\gamma$=7/5 (Goldreich \& Scoville 1976).  According to our tests, 
the effect of a varied adiabatic index (in comparison to a constant 
one) is only a few percent in the intensities, which is in agreement 
with the result of Groenewegen (1994).

Additional cooling is provided by emission from the molecules CO and
H$_2$.  The molecular cooling due to CO ($^{12}$CO and $^{13}$CO) is
calculated from the derived level populations using the expression of
Crosas \& Menten \cite*{crosas97}.  For H$_2$ we use the approach by
Groenewegen \cite*{groenewegen94}.  HCN could be an important coolant 
in the inner parts of the envelope (Cernicharo et al.  
1996)\nocite{cerni}, but Groenewegen \cite*{groenewegen94} 
demonstrates that HCN cooling is only of minor importance.  In 
the present version of the code HCN cooling is therefore not included.

The heating is assumed to be dominated by dust-gas collisions 
(Groenewegen 1994)\nocite{groenewegen94} and the photoelectric effect 
(Crosas \& Menten 1997)\nocite{crosas97}.  In the former case a number 
of free parameters describing the dust are introduced.  These are 
highly uncertain but affect the derived line intensities.  Here we 
assume that the $Q$-parameter, i.e., the efficiency of momentum 
transfer, to be equal to 0.03, and define a new parameter that 
contains the other dust parameters, $h=(\Psi/0.01) (2.0\, 
\mathrm{g\,cm}^{-3} / \rho_\mathrm{d}) (0.05\,\mu\mbox{m} / 
a_\mathrm{d})$, where $\Psi$ is the gas-to-dust ratio, 
$\rho_\mathrm{d}$ is the dust density, and $a_\mathrm{d}$ the mean 
size of the dust particles.  The normalised values are the ones we 
used to fit the CO line emission of \object{IRC+10\,216} using our 
model (see Sch\"oier et al.  in prep.).  Heating (or cooling) due to a 
temperature difference (heat exchange) between the dust and the gas 
could affect the derived temperature structure of the gas and hence 
the line intensities, especially in the inner parts where the coupling 
between dust and gas is high.  However, using the dust temperature 
structure for \object{IRAS 15194-5115} from Lopez et al.\ 
\cite*{lopez} and the analytical expression from Groenewegen 
\cite*{groenewegen94}, we conclude that this effect is only of minor 
importance.

\object{IRAS 15194-5115} is a LPV with a period of 580 days, and with
a variation of 0.8$^\mathrm{mag}$ in the Johnson M filter
\cite{lebertre}, but the amplitude decreases rapidly towards longer
wavelengths.  The period-luminosity relation for C-stars 
(Groenewegen \& Whitelock 1996\nocite{groen})
gives a luminosity of 8770\,L$_\odot$.
Using this luminosity and the apparent bolometric
flux we arrive at a distance of $600\,\mbox{pc}$, a value in agreement
with the ones discussed recently in the literature.  The stellar
temperature is set to $T_{\mathrm{eff}}=2200$\,$\mbox{K}$, a value typical
for C-stars with a high mass loss rate (e.g., Le Bertre 1997)\nocite{lebertre97}.
However, the mass loss rate derived is quite insensitive to the
choice of $T_{\mathrm{eff}}$, since radiative excitation is of very little
importance in this case.  The reason is that, as our calculations
show, in this high-density CSE the populations in the CO rotational
levels are determined by collisions, and hence the kinetic temperature,
out to a few$\times$10$^{16}$\,cm (Fig.~\ref{model}).  Consequently,
the light variability of the object is also unimportant for the
excitation.  In the present version of our code only the star
provides radiation since no thermal emission from dust in the CSE is
included.  This latter radiation could be significant for pumping vibrational
states, leading to non-thermal level populations of the rotational states
which could effect the FIR line intensities, but this effect is judged
to be small in this high-density CSE.

We assume a spherically symmetric CSE formed by a constant
mass loss rate and expanding at a constant velocity.  From the widths of
the radio lines we are able to estimate the terminal gas expansion
velocity, $\sim$21.5\,km\,s$^{-1}$.  The turbulent velocity width affects the
effectiveness of the radiative excitation.  However, since this is
only of minor importance, we will use a constant value of 0.5\,km\,s$^{-1}$
throughout the envelope.  The inner radius we set to 4\,R$_*$; the
stellar radius being estimated to be 4.5$\times$10$^{13}$\,cm [we note
here that Lopez et al.  \cite*{lopez} found the dust condensation
radius to be about 6\,R$_*$, where the dust temperature is
$\sim$$1000\,\mbox{K}$].  The inner parts of the CSE are complicated
regions of which we have a very limited understanding, see e.g.  Ryde
et al.  \cite*{ryde_CO2}, and where many of the basic assumptions of
our model are certainly not valid.  The properties of this region
could, in principle, be of importance for the excitation of the higher
rotational levels.  According to our tests, however, an initial
velocity gradient for example, affects the intensities only 
marginally.  We have also performed tests showing that a change of the 
inner radius by a factor of two affects the derived FIR line 
intensities by at most a few percent.

The photodissociation radius of the CO envelope is defined as the
location where the fractional abundance of CO is half the photospheric
value.  This was estimated to be 3.2$\times$\,$10^{17}$\,cm for
$^{12}$CO (for the estimated mass loss rate) using the calculations by
Mamon et al.  \cite*{mamon}.  The two isotopic variants have different
photodissociation and chemical fractionation behaviour, but according 
to Mamon et al.  \cite*{mamon} these effects tend to cancel out, and 
the authors predict that the $^{12}$CO/$^{13}$CO-ratio is constant 
with radius, except that the $^{13}$CO envelope extends slightly 
further out.  Here we assume the same radius for both isotopic 
variants.  The $J$=1--0 lines are affected the most when varying the 
outer radius, but according to our calculations only by $\sim$20\% in 
intensity, even if the outer radius is changed by a factor of two.  We 
adopt an initial abundance of $^{12}$CO with respect to H$_2$, $f_{\rm 
CO}$, of 10$^{-3}$.  The estimated mass loss rate scales roughly as $f_{\rm 
CO}^{-0.8}$.  The $^{13}$CO abundance is obtained from the best-fit 
model.

\section{Results and discussion}

When running our model we vary the free parameters, the mass loss 
rate, the $^{12}\mbox{CO}/^{13}\mbox{CO}$-ratio, and the dust 
parameter $h$, until we get a good fit to the observed $^{12}$CO and 
$^{13}$CO radio line intensities, profiles, and radial brightness 
distributions (Nyman et al.  1993\nocite{nyman93}).  This gives the 
kinetic temperature structure of the CSE, albeit constrained only by 
emission from the outer regions.  Finally, we compare the model fluxes 
for the higher rotational transitions with our LWS observations.  In 
Figs~\ref{radio} \& \ref{iso} we have plotted the line profiles 
(convolved with the instrumental resolutions) of the best-fit model to 
the $^{12}$CO lines (there are narrow interstellar CO lines present in 
the blue wing of the radio lines, and we exclude these lines when 
integrating the radio fluxes).  We also plot the best model fit to the 
$^{13}$CO lines for the best estimate of the 
$^{12}\mbox{CO}/^{13}\mbox{CO}$-ratio.  

From a simple error analysis, we arrive at a mass loss rate (based on 
the radio observations) of 
(1$\pm$0.2)$\times$10$^{-5}$\,\mbox{M$_\odot$\,yr$^{-1}$} (excluding 
uncertainties in distance, $^{12}\mbox{CO}$ abundance, and the assumed 
circumstellar model), in agreement with other estimates (Nyman et al.  
1993\nocite{nyman93}; Le Bertre 1997\nocite{lebertre97}).  
We obtain a $^{12}\mbox{CO}/^{13}\mbox{CO}$-ratio of 5.5$\pm$1.  This 
result agrees excellently with that found by Nyman et al.  
\cite*{nyman93} using intensity ratios of isotopic variants of CS, CN, 
HCN, and HC$_3$N.

\begin{figure*}
  \resizebox{\hsize}{!}{\includegraphics{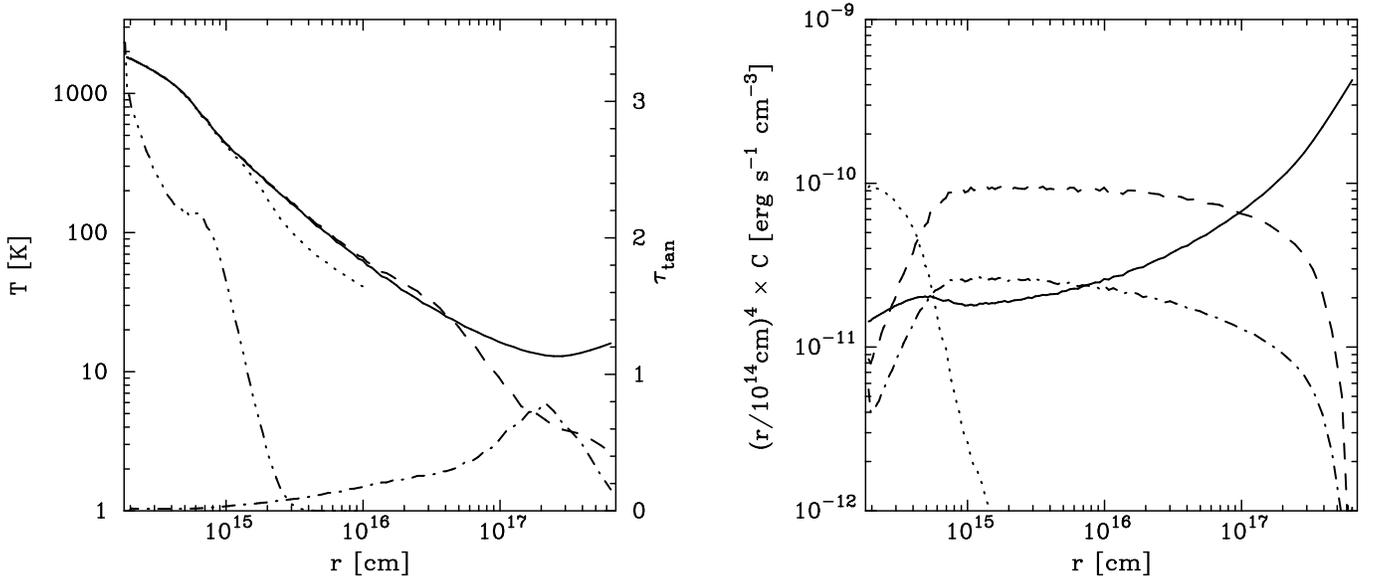}}
  \caption{The full line in the temperature panel gives the gas kinetic 
  temperature, the dashed [dotted] line is the excitation temperature of 
  the $^{12}$CO(1--0) [$^{12}$CO(18--17)] transition.  The dash-dot 
  [dash-dot-dot-dot] line gives the tangential optical depth, 
  $\tau_{\mbox{tan}}$, of the $^{12}$CO(1--0) [$^{12}$CO(18--17)] 
  transition.  In the cooling panel the full line represents adiabatic 
  cooling; the dotted line the H$_2$ cooling; and the dashed line 
  and the dash-dot line the $^{12}$CO and $^{13}$CO cooling, 
  respectively (see text for details)}
  \label{model}
\end{figure*}

The kinetic and excitation (for selected transitions) temperature 
structures are plotted in Fig.~\ref{model}.  The $^{12}$CO(1--0) 
transition is close to thermalised out to 
$\sim$3$\times$10$^{16}$\,cm.  The $^{13}$CO(1--0) transition is 
strongly super-thermal in the inner parts, and it is never 
thermalised.  The FIR lines are thermalised only out to about 
1$\times$10$^{15}$\,cm.  The radio lines are formed mainly beyond 
$\sim$5$\times$10$^{16}$\,cm ($\sim$1000$\mbox{R$_*$}$), corresponding 
to a time scale of $\ga$10$^3$\,yr, while the FIR lines are formed 
much closer to the star, within 10$^{15}$\,cm 
($\sim$10$\mbox{R$_*$}$), which corresponds to $\la$20\,yr.

In Fig.~\ref{model} we also plot the radial variation of the various
cooling terms included in the energy balance equation,
Eq.~(\ref{ebal}).  Here, for clarity, the cooling rate has been multiplied by a
factor $\propto r^4$ (the heating due to dust-gas collisions goes
roughly as $r^{-4}$).  CO cooling (mainly by
$^{12}$CO) is the dominant coolant in the region between
$\sim$4$\times$10$^{14}$\,cm and $\sim$1$\times$10$^{17}$\,cm, where
most of the emission from the observed transitions emanates.  In this
context a high abundance of $^{13}$CO could affect the
modelled line intensities significantly.  For \object{IRAS\,15194-5115} a
$^{12}\mbox{CO}/^{13}\mbox{CO}$-ratio of only 5.5 implies that cooling by
$^{13}$CO could be as high as 20\% of that of $^{12}$CO.  We have
inluded $^{13}$CO cooling in the same self-consistent way as for
$^{12}$CO and it turns out that its relative cooling is somewhat higher
than 20\%, Fig.~\ref{model}, due to the fact that the lines of $^{13}$CO 
are optically thinner than those of $^{12}$CO.  Closer to the star H$_2$
cooling dominates but very little emission from this part of the
envelope contributes to the line emission observed.  In the cool,
tenuous, outer parts of the envelope, where most of the observed 
\mbox{CO(1--0)} emission arises, cooling due to the adiabatic 
expansion dominates.

We derive a dust parameter $h$$\sim$2 (with an estimated error of
about 30\%), i.e.  twice as large as the one we used to fit the CO
radio lines of \object{IRC+10\,216} in our modelling of this object.
This indicates that the dust properties or the relative amount of dust
could be different in the two objects.  Our result is consistent with
the finding by Groenewegen et al.  \cite*{groen98} in their dust models
that \object{IRAS\,15194-5115} has a dust-to-gas ratio which is about
twice the value for \object{IRC+10\,216} (using the same values for
other dust parameters).  However, our derived dust parameter is sensitive to
the adopted luminosity, and hence the result is uncertain.

The model fluxes compare well with the observed FIR fluxes within the 
estimated errors, see Table~1, although we note that the fluxes from 
lines at longer wavelengths are systematically higher compared to what 
the model predicts, possibly indicating a systematic error.  We 
conclude that the FIR data, probing the inner regions of the CSE, are 
consistent with the model derived from radio observations.  
Thus, we suggest that there have been no significant changes in the 
stellar wind characteristics over the past $\sim$5000 years.  Due to 
the estimated large uncertainties in the LWS data the mass loss rate 
in the inner parts may have changed by about a factor of three and 
still be consistent with the model.  Likewise, there appears to have 
been no significant change in the $^{12}\mbox{CO}/^{13}\mbox{CO}$-ratio 
over this period.

We believe that the circumstellar $^{12}\mbox{CO}/^{13}\mbox{CO}$-ratio 
is very close to the $^{12}\mbox{C}/^{13}\mbox{C}$-ratio in the 
stellar atmosphere.  A $^{12}\mbox{C}/^{13}\mbox{C}$-ratio of only 5.5 is 
just slightly larger than the equilibrium value of 3--3.5 found from hydrogen 
burning through the CNO-cycle.  Such a low 
$^{12}\mbox{C}/^{13}\mbox{C}$-ratio puts \object{IRAS\,15194-5115} 
close to the group of C-stars known as the J-type stars, which have 
$^{12}\mbox{C}/^{13}\mbox{C}$-ratios of 3--4 (Lambert et al.\ 
1986)\nocite{lambert}.  Optically bright C-stars often have ratios 
above 30 \cite{lambert}, and this seems to apply also for high mass 
loss rate objects (Greaves \& Holland 1997\nocite{greaves97}).  The 
best example of the latter is \object{IRC+10\,216} for which Kahane et 
al.  \cite*{kahane} derived a value of 
$^{12}\mbox{C}/^{13}\mbox{C}$=44$\pm$3.  From its 
$^{12}\mbox{C}/^{13}\mbox{C}$-ratio \object{IRAS\,15194-5115} 
resembles more closely the extreme OH/IR-stars which are massive AGB 
stars in the very last phases of evolution (Delfosse et al.  
1997\nocite{delfosse}).   The low $^{12}\mbox{C}/^{13}\mbox{C}$-ratio 
of the OH/IR stars may be caused by hot bottom burning.  This occurs 
when the temperature is larger than $\sim$80$\times 10^{6}\mbox{K}$ 
and this is only achieved in massive (5-8 M$_\odot$) AGB stars 
\cite{boot}.  The luminosity derived for \object{IRAS\,15194-5115} 
from the period-luminosity relation ($8770$ L$_\odot$) is, however, 
lower than typical OH/IR luminosities ($\sim$3$\times 10^{4}$ 
L$_\odot$).  Thus, \object{IRAS\,15194-5115} may have
been a J-star for which the $^{12}\mbox{C}/^{13}\mbox{C}$-ratio has
remained low despite it probably being highly evolved.  Alternatively,
it may be a fairly massive star, which has recently become a carbon
star due to quenching of hot bottom burning (Frost et al.
1998\nocite{frost98}).

We have shown that modelling radio and FIR lines simultaneously is potentially 
an interesting diagnostic method for probing the stellar wind history of giant stars. Of special interest are, e.g., the group of 
OH/IR stars which are suggested to be undergoing an onset of a superwind phase 
with a drastically enhanced mass-loss rate during the recent 500-1000 years 
(Delfosse et al.  1997\nocite{delfosse}).  This hypothesis could 
be observationally confirmed either with interferometric CO 
observations, as suggested by Delfosse et al. \cite*{delfosse} or by 
FIR observations of lines from rotationally excited CO molecules as 
is demonstrated in this paper.

\begin{acknowledgements}
We would like to thank L-\AA\, Nyman for providing us with the radio 
observations, F.  Kerschbaum and B.  Larsson for discussions, and K.  
Eriksson and B.  Gustafsson for providing valuable comments.  E.  
Olsson is thanked for generous assistance.  Support from the ISO 
Spectrometer Data Centre at MPE Garching, funded by DARA under grant 
50 QI 9402 3 is acknowledged.  The ISO Spectral Analysis Package 
(ISAP) is a joint development by the LWS and SWS Instrument Teams and 
Data Centers.  Contributing institutes are CESR, IAS, IPAC, MPE, RAL 
and SRON.
\end{acknowledgements}


\begin{thebibliography}{}

\bibitem[\protect\astroncite{{Boothroyd} et~al.}{1993}]{boot}
{Boothroyd} A.~I., {Sackmann} I.~J., {Ahern} S.~C., 1993,
\newblock { ApJ} { 416}, 762

\bibitem[\protect\astroncite{{Cernicharo} et~al.}{1996}]{cerni}
{Cernicharo} J., {Barlow} M., {Gonzalez-Alfonso} E., et~al., 1996,
\newblock { A\&A} { 315}, L201

\bibitem[\protect\astroncite{{Chandra} et~al.}{1996}]{chandra}
{Chandra} S., {Maheshwari} V., {Sharma} A., 1996,
\newblock { A\&AS} { 117}, 557

\bibitem[\protect\astroncite{{Clegg} et~al.}{1996}]{LWS}
{Clegg} P., {Ade} P., {Armand} C., et~al., 1996,
\newblock { A\&A} { 315}, L38

\bibitem[\protect\astroncite{{Crosas} and {Menten}}{1997}]{crosas97}
{Crosas} M., {Menten} K.~M., 1997,
\newblock { ApJ} { 483}, 913

\bibitem[\protect\astroncite{{Delfosse} et~al.}{1997}]{delfosse}
{Delfosse} X., {Kahane} C., {Forveille} T., 1997,
\newblock { A\&A} { 320}, 249

\bibitem[\protect\astroncite{{Epchtein} et~al.}{1987}]{epstein}
{Epchtein} N., {Le Bertre} T., {Lepine} J. R.~D., et~al., 1987,
\newblock { A\&AS}, { 71}, 39

\bibitem[\protect\astroncite{{Flower} and {Launay}}{1985}]{flower}
{Flower} D.~R.,  {Launay} J.~M., 1985,
\newblock { MNRAS} { 214}, 271

\bibitem[\protect\astroncite{{Frost} et~al.}{1998}]{frost98}
{Frost} C.~A., {Cannon} R.~C., {Lattanzio} J.~C., {Wood} P.~R.,
  {Forestini} M., 1998,
\newblock { A\&A} { 332}, L17

\bibitem[\protect\astroncite{{Goldreich} and {Scoville}}{1976}]{goldreich76}
{Goldreich} P., {Scoville} N., 1976,
\newblock { ApJ} { 205}, 144

\bibitem[\protect\astroncite{{Greaves} and {Holland}}{1997}]{greaves97}
{Greaves} J.~S., {Holland} W.~S., 1997,
\newblock { A\&A} { 327}, 342

\bibitem[\protect\astroncite{{Groenewegen}}{1994}]{groenewegen94}
{Groenewegen} M. A.~T., 1994,
\newblock { A\&A} { 290}, 531

\bibitem[\protect\astroncite{{Groenewegen} and {Whitelock}}{1996}]{groen}
{Groenewegen} M. A.~T., {Whitelock} P.~A., 1996,
\newblock { MNRAS} { 281}, 1347

\bibitem[\protect\astroncite{{Groenewegen} et~al.}{1998}]{groen98}
{Groenewegen} M. A.~T., {Whitelock} P.~A., {Smith} C.~H., {Kerschbaum}
  F., 1998,
\newblock { MNRAS} { 293}, 18

\bibitem[\protect\astroncite{{Kahane} et~al.}{1992}]{kahane}
{Kahane} C., {Cernicharo} J., {Gomez-Gonzalez} J., {Guelin} M., 1992,
\newblock { A\&A} { 256}, 235

\bibitem[\protect\astroncite{{Kessler} et~al.}{1996}]{kessler}
{Kessler} M.~F., {Steinz} J.~A., {Anderegg} M.~E., et~al., 1996,
\newblock { A\&A} { 315}, L27

\bibitem[\protect\astroncite{{Lambert} et~al.}{1986}]{lambert}
{Lambert} D.~L., {Gustafsson} B., {Eriksson} K., {Hinkle} K.~H., 1986,
\newblock { ApJS} { 62}, 373

\bibitem[\protect\astroncite{{Le Bertre}}{1992}]{lebertre}
{Le Bertre} T., 1992,
\newblock { A\&AS} { 94}, 377

\bibitem[\protect\astroncite{{Le Bertre}}{1997}]{lebertre97}
{Le Bertre} T., 1997,
\newblock { A\&A} { 324}, 1059

\bibitem[\protect\astroncite{{Lopez} et~al.}{1993}]{lopez}
{Lopez} B., {Perrier} C., {Mekarnia} D., {Lefevre} J., {Gay} J., 1993,
\newblock { A\&A} { 270}, 462

\bibitem[\protect\astroncite{{Mamon} et~al.}{1988}]{mamon}
{Mamon} G.~A., {Glassgold} A.~E., {Huggins} P.~J., 1988,
\newblock { ApJ} { 328}, 797

\bibitem[\protect\astroncite{{Meadows} et~al.}{1987}]{meadows}
{Meadows} P.~J., {Good} A.~R., {Wolstencroft} R.~D., 1987,
\newblock { MNRAS} { 225}, 43P

\bibitem[\protect\astroncite{{Nyman} et~al.}{1993}]{nyman93}
{Nyman} L.-\AA ., {Olofsson} H., {Johansson} L. E.~B., et~al., 1993,
\newblock { A\&A} { 269}, 377

\bibitem[\protect\astroncite{{Olofsson}}{1996}]{olofsson96}
{Olofsson} H., 1996,
\newblock { Ap\&SS} { 245}, 169

\bibitem[\protect\astroncite{{Ryde} et~al.}{1999}]{ryde_CO2}
{Ryde} N., {Eriksson} K., {Gustafsson} B., 1999,
\newblock { A\&A} { 341}, 579

\bibitem[\protect\astroncite{{Swinyard} et~al.}{1996}]{swinyard}
{Swinyard} B., {Clegg} P., {Ade} P., et~al., 1996,
\newblock { A\&A} { 315}, L43

\end{thebibliography}

\end{document}